# Quantum mechanical comparison between lithiated and sodiated silicon nanowires

**Donald C. Boone***

Nanoscience Research Institute, USA.
*Correspondence: db2585@caa.columbia.edu

**Abstract:** This computational research study will compare the specific charge capacity (SCC) between lithium ions inserted into crystallize silicon (c-Si) nanowires versus sodium ions inserted into amorphous silicon (a-Si) nanowires. It will be demonstrated that the potential energy V(r) within the lithium-silicon nanowire supports a coherent energy state model with discrete electron particles while the sodium-silicon nanowire potential energy will be discovered to be essentially zero and thus the electron current that travels through the sodiated silicon nanowire will be modeled as free electron with wave-like characteristics. This is due to the vast differences in the electric fields of the lithiated and sodiated silicon nanowires where the electric fields are of the order of $\mathbf{10^{10}}$ V/m and $\mathbf{10^{-15}}$ V/m respectively. The main reason for the great disparity in electric fields are due to the present of optical amplification within lithium ions and the absence of this process within sodium ions. It will be shown that optical amplification develops coherent optical interactions which is the primary reason for the surge of specific charge capacity in the lithiated silicon nanowire. Conversely, the lack of optical amplification is the reason for the incoherent optical interactions within sodium ions which is the reason for the low presence of SCC in sodiated silicon nanowires.

## 1. Introduction

For over a decade lithium ion batteries (LIBs) has been a commercial success for consumer electronics and electric vehicles. However lithium is a relatively strategic and precious metal that will be limited due to the increase use by developing countries as we advance into the 21st century. Sodium ion (Na+) batteries have created attention due to Na abundance worldwide. Similar to the lithium atom, Na is a Rydberg atom (one electron in the outer atomic orbit) although the sodium atom is larger than lithium [1].

One of the original commercial LIB designs in the 1990s used carbon base anode material for varies energy storage devices with a specific charge capacity (SCC) of approximately 400 mA-hr/g [2]. In order to improve the SCC of lithium ion batteries, crystallize silicon (c-Si) has been studied as a possible substitute anode material instead of carbon. The SCC of lithiated silicon has been found by several research studies to be over 4000 mA-hr/g which made c-Si anode material for lithium ions very popular in LIB research [3]. Unfortunately, as was discovered during the course of this research that lithium ion insertion in crystallize silicon produces extremely large anisotropic volume expansion in the range of 300 to 400 percent of the lithiated silicon nanowire original volume [4]. This volume expansion has led to an overall decrease in the specific charge capacity and ultimately the failure of this LIB component [5]. This research that has been studied using MD/DFT simulations [6] has uncover problems that has not been overcome in the design and manufacturing of lithiated silicon batteries for commercial use.

Concurrently with the research of lithium ion insertion into crystallizes silicon, the study of sodium ion insertion into the same anode material of c-Si was being performed.

In this research the formation energy of sodium-silicon $Na_xSi$ was found to be significantly lower than that of lithium-silicon $Li_xSi$ (where x is the ratio of lithium or sodium ions to silicon atoms) [7]. This caused the crystallized silicon to be nonreactive to sodium ions and therefore a poor anode material [8]. During this time amorphous silicon (a-Si) was substituted for crystallize silicon (c-Si) in sodium ion/silicon research. It was discovered that a-Si had a formation energy substantially higher than c-Si when sodium ions were inserted and therefore amorphous silicon could possibly be used as an anode material for batteries [9]. As a result this research will be comparing the reactions between $Li^+$/c-Si and $Na^+$/a-Si materials.

Two in-situ theoretical apparatuses will be the focus of this research, one for a lithium ion/crystallized silicon (c-Si) nanowire and the other a sodium ion/amorphous silicon (a-Si) nanowire as displayed in figure 1a and 1b. Prior to the beginning of the lithiation of the c-Si nanowire and sodiation of the a-Si nanowire, the individual lithium and sodium atoms are ionized reducing them to their constitutive particles of lithium/sodium ions and free electrons. Each silicon nanowire is part of a separate electric series circuit with a constant 2-V voltage that is applied to each nanowire in order to diffuse lithium and sodium ions through their respective nanowire. In both lithiated and sodiated silicon nanowires, the electrons and lithium/sodium ions enter the c-Si and a-Si silicon nanowire respectively at opposing ends and therefore travel in opposite directions. Since the electrons and ions are moving charge particles, they are the source of the electric fields. The c-Si nanowire is compose of diamond crystalline cubic structure as shown in figure 1c whereas the a-Si nanowire is comprised of the same number of silicon atoms albeit in a random formation.

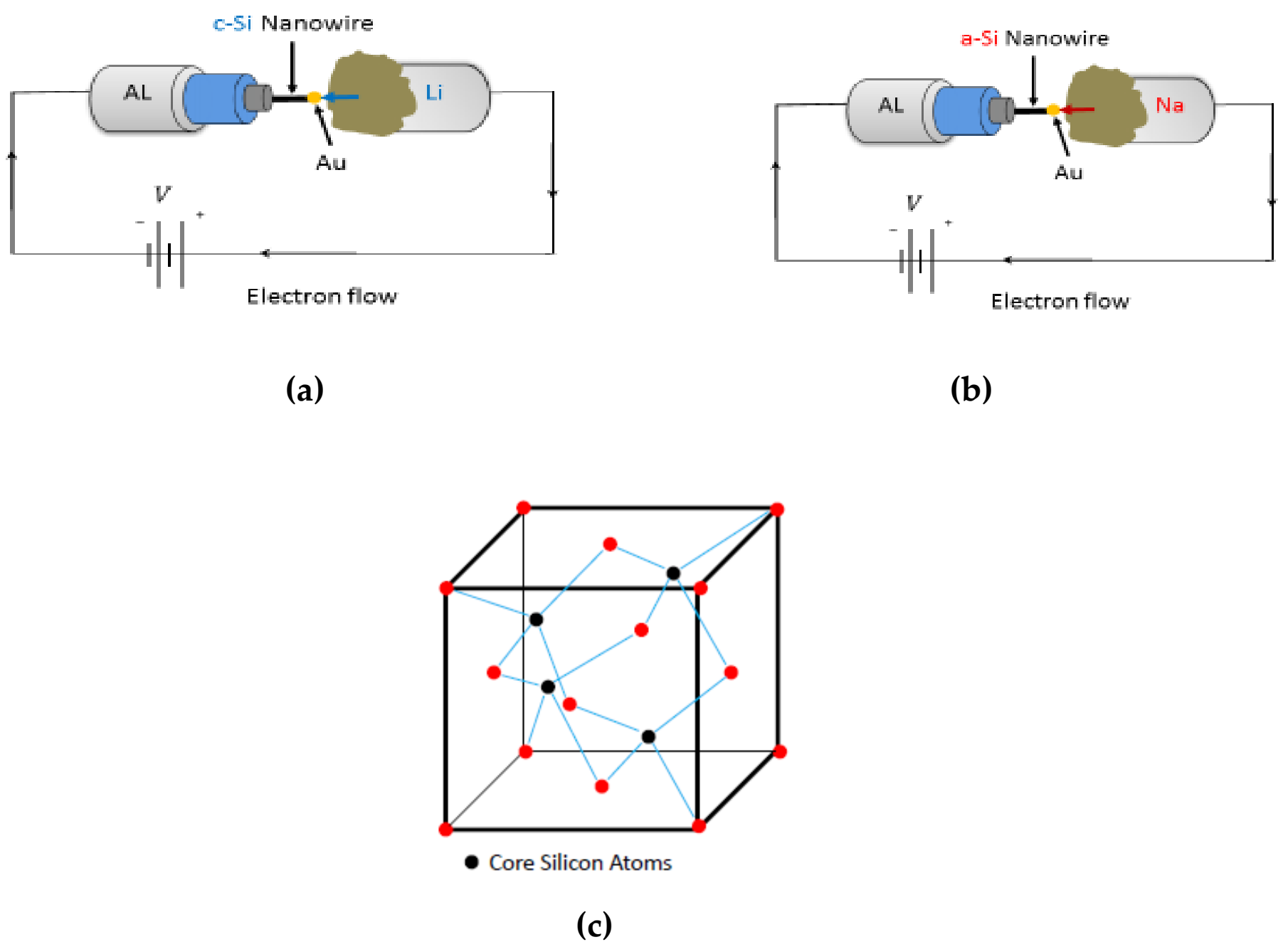

**Figure 1.** In-situ apparatus arrangement for a solid electrochemical cell using (**a**) lithium metal counter electrode (Li) and (**b**) sodium metal counter electrode (Na) each with an applied voltage 2-V source. (c) Silicon is a diamond crystalline cubic structure made up of tetrahedral molecules with its hybridized $sp^3$ orbitals within their valence shells filled with covalent bonding electrons from neighboring silicon atoms.

## 2. Electric Fields

In order to define the induced electric fields for lithiated and sodiated silicon nanowires, the wavefunctions must first be derived for each of the constitutive particles presented in this study. The ion diffusion and electron current within the silicon nanowires are assumed to be continuous uniform flow with minimal temporal variations. For this reason the time-independent Schrodinger equation will be used to solve the wavefunctions for lithium, sodium and silicon.

$$\left[-\frac{\bar{h}^2}{2m_{eff}}\nabla^2 + V(r)\right]\Psi_B(r) = E\Psi_B(r) \quad (1)$$

where $\bar{h}$ is the Planck's constant, $m_{eff}$ is the effective mass of the electron, E is the eigenvalues and V(r) is the potential energy. The ground state wavefunctions that will be used for lithium ions and silicon atoms are

$$\Psi_B(r,\theta,\phi) = Nr^{n-1}\exp[Z_{eff}\left(\frac{\bar{a}_o}{n}\right)]\, Y_l^m(\theta,\phi) \quad (2a)$$

$$\bar{a}_o = -\frac{r}{a_o} \quad (2b)$$

where $r, \theta, \phi$ are the spherical coordinates, n the energy level, $Z_{eff}$ is the effective atomic number, N is the normalization constant, $a_o$ is Bohr radius and $Y_l^m(\theta,\phi)$ is the spherical harmonics. The subscript B denotes the type of wavefunction that will be used which will be lithium ions ($Li^+$) and silicon atoms (c-Si, a-Si). The wavefunctions $\Psi_{Li}$, $\Psi_{c_Si}$ and $\Psi_{a_Si}$ are calculated using the Slater determinant. The ground state wavefunction for sodium ions $\Psi_{Na}$ was defined by using a harmonic oscillator wavefunction

$$\Psi_{Na} = Ae^{-\frac{m_{eff}\,\omega}{2\bar{h}}r^2} \quad (3)$$

where A is the amplitude and ω is the angular frequency of the electron. In addition to equation 3 being a solution to the time-independent Schrodinger equation it is also a solution to the Thomas-Fermi equation which is part of a theoretical model for the electronic structure of atoms that is the predecessor to density functional theory [10]

$$\frac{d^2\Psi_{Na}}{dx^2} = \frac{{\Psi_{Na}}^{3/2}}{x^{1/2}} \quad (4)$$

where $x=0.885Z_{eff}^{-\frac{1}{3}}\, a_o r$.

The ground state wavefunctions are embedded within the electric field equations via the Bloch function equations. This is accomplished by defining the expectation value of the wave numbers of lithium ions, sodium ions and silicon atoms. First, the wave number expectation values for lithium ions and crystallized silicon atoms are

$$k_{Li} = \frac{\langle\Psi_{Li}|\hat{k}|\Psi_{Li}\rangle}{\langle\Psi_{Li}|\Psi_{Li}\rangle} \qquad k_{c_Si} = \frac{\langle\Psi_{c_Si}|\hat{k}|\Psi_{c_Si}\rangle}{\langle\Psi_{c_Si}|\Psi_{c_Si}\rangle} \quad (5a,b)$$

where $\hat{k} = -i\nabla$. The Bloch functions of lithium ion $u_{Li}(r)$ and crystallized silicon atom $u_{c_Si}(r)$ are defined as

$$u_{Li}(r) = e^{ik_{Li}r} + \frac{1}{k_{Li}r}e^{i(\delta_{Li}+k_{Li}r)}\sin\delta_{Li} + \frac{3z}{k_{Li}r^2}e^{i(\delta_{Li}+k_{Li}r)}\sin\delta_{Li} \quad (6)$$

$$u_{c_Si}(r) = e^{ik_{c_Si}r} + \frac{1}{k_{c_Si}r}e^{i(\delta_{c_Si}+k_{c_Si}r)}\sin\delta_{c_Si} + \frac{3z}{k_{c_Si}r^2}e^{i(\delta_{c_Si}+k_{c_Si}r)}\sin\delta_{c_Si} \quad (7)$$

where $\delta_{Li}$ and $\delta_{c_Si}$ are defined as the phase shift of lithium ion and crystallized silicon atom respectively. The Bloch functions are in turn a function of the electric field $\vec{E}_{Li}$ utilizing the Drude model for electron transport within the $Li^+$ and c-Si matrix.

$$\bar{E}_{Li} = iC_E\frac{\bar{h}^2(3\pi^2\bar{n}_c)^{\frac{2}{3}}v_{DOS}}{4n_v e m_{eff}}\left[u_{c_Si}\nabla u^*_{Li} - u^*_{c_Si}\nabla u_{Li}\right] \quad (8)$$

As mentioned previously, negative free electrons and positive lithium ions enter into silicon nanowire model from opposing directions. The electric charge difference between the constitutive particles, where the electrons are always greater or equal in number to the lithium ions in the model, will be known as the average negative charge differential $\bar{n}_c$ which are the number of charge particles per unit volume. The electric charge unit of an electron is denoted as e, $n_v$ is the electron density of the maximum valence band, $v_{DOS}$ is defined as the density of state volume and coefficients $C_E$ allows the electric field to be a solution to Maxwell equations. The electric field $\vec{E}_{Li}$ describes the electrons in the minimum conduction band of lithium ions. These electrons are in the lowest energy state in the conduction band, however when the majority of lithium ions enter into the excited state, population inversion occurs and the electric field increases in strength through optical amplification factor $\gamma_{Li}$. As a result the lithium ion electric field $\vec{\mathbf{E}}_{\mathbf{Li}}$ is re-defined as

$$\vec{\mathbf{E}}_{\mathbf{Li}} = \vec{E}_{Li} \exp^{\frac{\gamma_{Li} r}{2}} \tag{9}$$

where

$$\gamma_{Li}(r,t) = \sigma_{Li} \cdot \Delta N_{Li} \tag{10}$$

$$\Delta N_{Li} = (N_{Li} - \vec{N}_1) \tag{11}$$

$$\sigma_{Li} = A_{Li} \frac{L^2}{8\pi\, n_{Li}^2} g(\omega) \tag{12}$$

In Equation (11) $\Delta N_{Li}$ is the difference of the number of excited state lithium ions $N_{Li}$ and the number of ground state silicon atoms $N_1$ within the diamond cubic lattice. The reason the silicon atoms are modeled to be in the ground state is because of their low electron transition probability in silicon atoms due to them being an indirect band gap material [11]. Therefore in this research study, $N_{Li}$ = 30 and $N_1$ =8 for a ratio of x = $N_{Li}/N_1$=3.75 which is the same value of x in lithiated silicon $Li_xSi$ and at which the silicon diamond cubic lattice in our model is considered to be at full lithiation. Equation (12) is the stimulated emission cross section area $\sigma_{Li}$ which is defined by the Einstein A Coefficient $A_{Li}$, the spectral line shape function $g(\omega)$, wavelength of the photon emitted L and the lithium refractive index $n_{Li}$. The total electric field becomes

$$\vec{\mathbf{E}}_{\mathbf{Li}} = iC_E \frac{\hbar^2 (3\pi^2 \bar{n}_c)^{\frac{2}{3}} v_{DOS}}{4 n_v e m_{eff}} \left[u_{c_Si} \nabla u^*_{Li} - u^*_{c_Si} \nabla u_{Li}\right] \exp^{\frac{\gamma_{Li} r}{2}} \tag{13}$$

A similar derivation for the electric field $\vec{\mathbf{E}}_{\mathbf{Na}}$ that is generated by the flow of electron current and the insertion of sodium ions ($Na^+$) into amorphous silicon (a-Si) is:

$$k_{Na} = \frac{\langle \Psi_{Na} | \hat{k} | \Psi_{Na} \rangle}{\langle \Psi_{Na} | \Psi_{Na} \rangle} \qquad k_{a_Si} = \frac{\langle \Psi_{a_Si} | \hat{k} | \Psi_{a_Si} \rangle}{\langle \Psi_{a_Si} | \Psi_{a_Si} \rangle} \tag{14a,b}$$

$$u_{Na}(r) = e^{ik_{Na} r} + \frac{1}{k_{Na} r} e^{i(\delta_{Na} + k_{Na} r)} \sin \delta_{Na} + \frac{3z}{k_{Na} r^2} e^{i(\delta_{Na} + k_{Na} r)} \sin \delta_{Na} \tag{15}$$

$$u_{a_Si}(r) = e^{ik_{a_Si} r} + \frac{1}{k_{a_Si} r} e^{i(\delta_{a_Si} + k_{a_Si} r)} \sin \delta_{a_Si} + \frac{3z}{k_{a_Si} r^2} e^{i(\delta_{a_Si} + k_{a_Si} r)} \sin \delta_{a_Si} \tag{16}$$

$$\gamma_{Na}(r,t) = \sigma_{Na} \cdot \Delta N_{Na} \tag{17}$$

$$\Delta N_{Na} = (N_{Na} - N_1) \tag{18}$$

$$\sigma_{Na} = A_{Na}\frac{\lambda^2}{8\pi n_{Na}^2}g(\omega) \tag{19}$$

$$\vec{\mathbf{E}}_{\mathbf{Na}} = iC_E\frac{\bar{h}^2(3\pi^2\bar{n}_c)^{\frac{2}{3}}v_{DOS}}{4n_v e m_{eff}}\left[u_{a_Si}\nabla u^*_{Na} - u^*_{a_Si}\nabla u_{Na}\right]\exp^{\frac{\gamma_{Na}r}{2}} \tag{20}$$

For the sodium ions/amorphous silicon nanowire model the full sodiation is at x=0.75 for sodiated silicon $Na_xSi$. This translate into $N_{Na} = 6$ for the excited state $Na^+$ions and $N_1$=8 for the ground state a-Si atoms for a ratio of x = $N_{Na}/N_1$. As stated previously, in order for optical amplification $\gamma$ to occur at least half the atoms/ions in the system must be in the excited state. This does not happen since x is less than one. Therefore, optical amplification does not occur and $\gamma_{Na} \approx 0$ for sodium ions.

Comparing the two electric fields of $\vec{E}_{Li}$ and $\vec{E}_{Na}$ that are generated by lithium ions and sodium ions as they are inserted into their respective silicon nanowires are displayed in figure 2 and in figure 3. Both electric fields are generated by opposing electron current traveling counter to lithium ion diffusion in figure 1a and sodium ion diffusion as depicted figure 1b. The electron currents (where the silicon lattice constant is *a*) are defined as

$$I_{Li} = \frac{2\bar{n}_c{}^{1/3}a^2e^2\vec{\mathbf{E}}_{\mathbf{Li}}}{\bar{h}(3\pi^2)^{2/3}} \qquad I_{Na} = \frac{2^{1/2}ea^{1/3}\omega_{ew}}{(3\pi^2)^{1/3}\ \bar{N}_c} \tag{21a,b}$$

where

$$\omega_{ew} = -\frac{3\bar{n}_c a^3\bar{h}\ k_{a-Si}}{m_{eff}}\nabla\psi_{ew} \tag{22}$$

$$\psi_{ew} = N^{\frac{1}{2}}\left(\frac{2}{\pi}\right)^{\frac{1}{4}}e^{-\left(\frac{r}{a}\right)^2}e^{ikr} \tag{23}$$

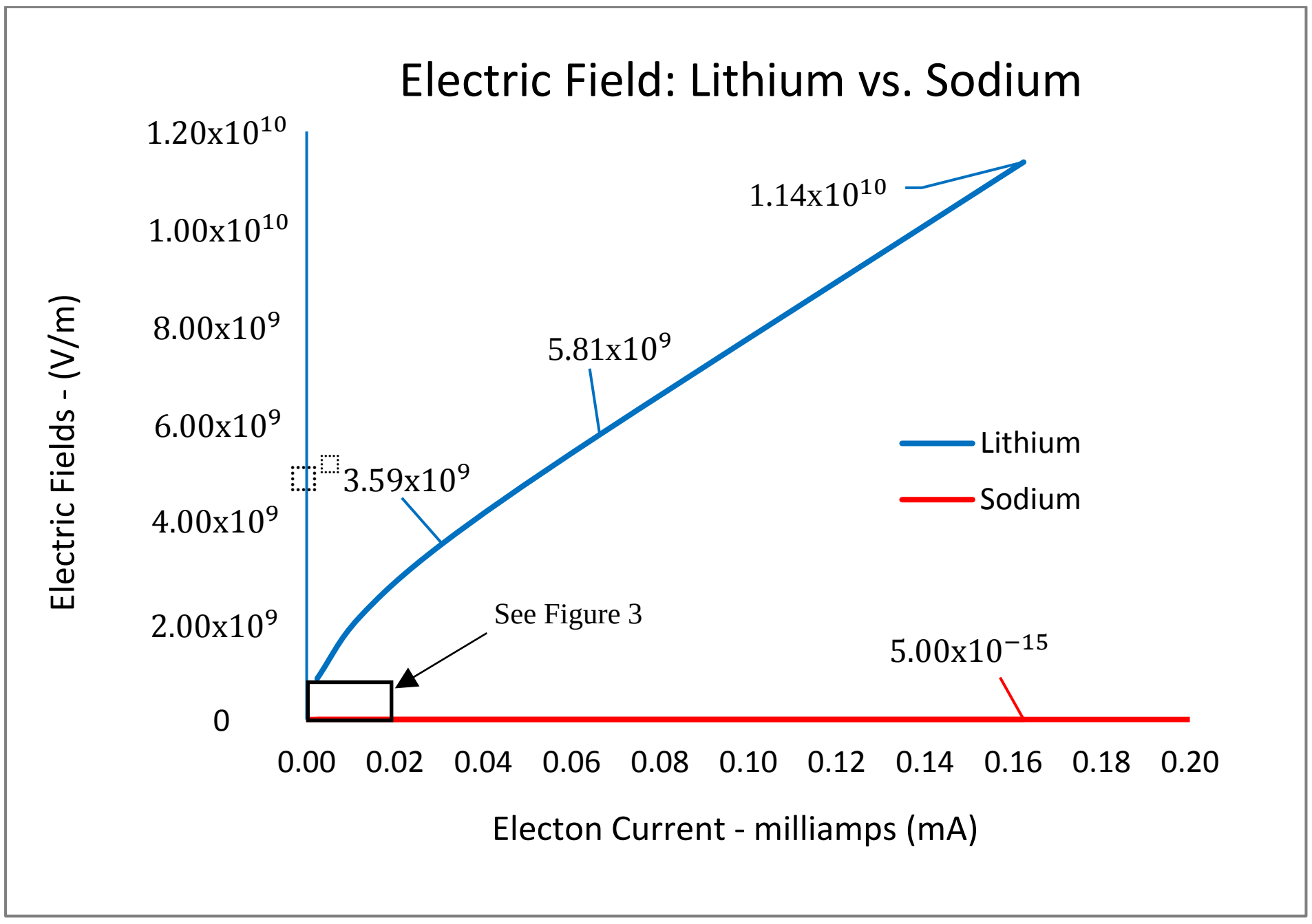

**Figure 2.** The electric fields that are generated within the lithiated and sodiated silicon nanowires respectively. The lithium ion electric field $\vec{E}_{Li}$ is of a magnitude between $10^9$ to $10^{10}$ V/m whereas the sodium ion electric field $\vec{E}_{Na}$ is of a magnitude $10^{-15}$V/m and therefore is approximately zero ($\vec{E}_{Na} \approx 0$).

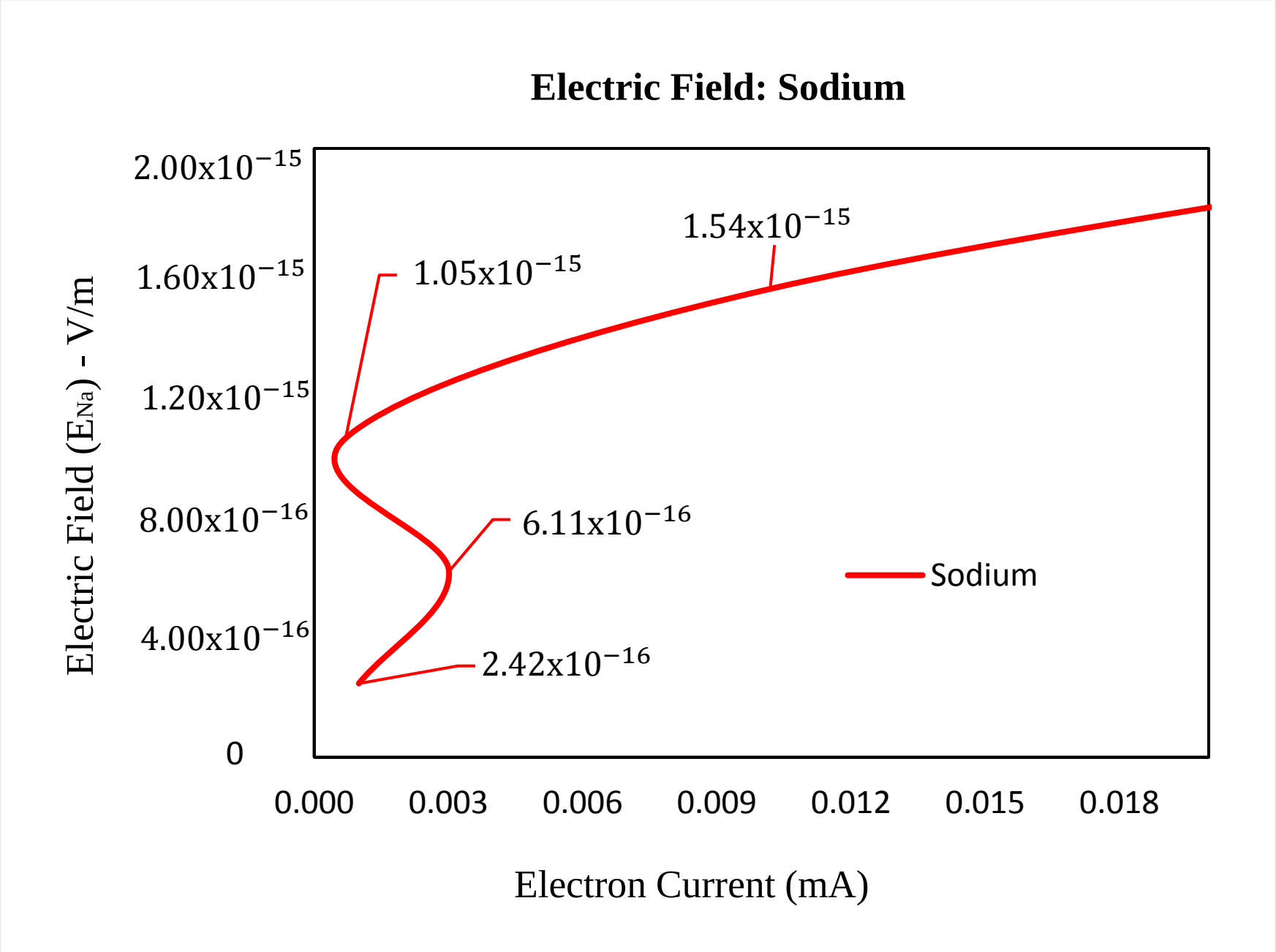

**Figure 3.** The sodium ion electric field $\vec{E}_{Na}$ has a wavelike characteristic at the quantum level and is approximately zero. Since the potential energy $V(r)$is a function of $\vec{E}_{Na}$, the electrons that travel within the $Na^{+}$/a-Si nanowire can be modeled as a quantum electron wave that is subjected to a potential energy of $V(r) = 0$.

The sodium electron current $I_{Na}$ is defined by the electron wave angular frequency $\omega_{ew}$ in which it is a function of the electron wavefunction $\psi_{ew}$ [12]. The number of electrons is represented by $\overline{N}_c$. The electric field $\vec{E}_{Li}$ in the Li/c-Si nanowire computational model is of a magnitude between $10^9$ to $10^{10}$ volt/meters compared to electric field $\vec{E}_{Na}$ in the $Na^{+}$/a-Si nanowire model is $10^{-16}$ to $10^{-15}$V/m. The potential energy $V(r)$ within the Hamiltonian representing both models are directly proportional to the electric fields. Since the electric field is greater in the lithiated silicon then the sodiated silicon nanowire, the energy states within each nanowire will be modeled differently based on quantum mechanical theory. For the $Li^{+}$/c-Si model, discrete energy states of a quantum harmonic oscillator will be utilized for the electron current and lithium ions within the c-Si nanowire. However, since the potential energy $V(r)$ in $Na^{+}$/a-Si model is approximately zero, the electron current will be modeled as free electrons that are unrestricted from the potential energy as they travel through the a-Si nanowire. This gives the electron current within the $Na^{+}$/a-Si nanowire model wavelike characteristics as oppose to the electron current within the $Li^{+}$/c-Si nanowire where the electrons exist not as waves but as a particles and only at discrete energy levels (eqs. 21-23).

## 3. Degree of Coherence

The electric field in each silicon nanowire will be analyzed in terms of their quantum optical interactions. These interactions are between the photons in the lithiated silicon nanowires and between the electric waves in the sodiated silicon nanowire. These quantum optical interactions are described by three types of interferences: coherent, incoherent and mixed interactions. From these interferences a set of phase matching conditions will be established:

Phase Matching Conditions:

1) Coherent Optical Interactions: $\omega_c = \sum_{i=1}^{M} \omega_i = M\omega$
Constructive Interference where $\omega_i$ are equal and are in-phase

2) Incoherent Optical Interactions: $\omega_c = \sum_{i=1}^{M} \omega_i \approx 0$
Destructive Interference where $\omega_i$ are not equal and not in-phase

3) Mixed Optical Interactions: $\omega_c = \sum_{i=1}^{M} \omega_i \neq 0$
Partial Destructive Interference is a combination of coherent and incoherent interactions.

M is the total number of particles (for lithium) or wave amplitudes (for sodium) for each interaction and $\omega_c$ is called the coherent angular frequency that will be defined in equations 26 and 32.

From the phase matching conditions the coherent optical states for each electric field will be the focus in determining the specific charge capacity (SCC) for the lithiated and sodiated silicon nanowires with respect to the electron current. The second-order correlation function $g^{(2)}$ will be used to calculate the degree of coherence, which in this research study is defined as the measure of the amount of coherent interference for the electric fields. For the lithium ion electric field the $g_{Li}^{(2)}$ is define as [13]

$$g_{Li}^{(2)} = 1 + \exp[-\pi(\alpha_{Li})^2] \tag{24}$$

where

$$\alpha_{Li} = \frac{\omega_{c_Li}}{\omega_{\gamma_Li}} \tag{25}$$

$$\omega_{c_Li} = \omega_{e_Li}\lambda_{Li} \tag{26}$$

$$\omega_{\gamma_Li} = \frac{a^3}{2\bar{h}}\left[ {n_{Li}}^2 {\vec{\mathbf{E}}_{\mathbf{Li}}}^2 \right] \tag{27}$$

$$\omega_{e_Li} = \frac{e\vec{\mathbf{E}}_{\mathbf{Li}}}{\bar{h}\,(3\pi^2\bar{n}_c)^{\frac{1}{3}}} \tag{28}$$

$$\lambda_{Li} = \frac{a^2 {m_{eff}}^{\frac{3}{2}} \omega_{e_{Li}}^{\frac{3}{2}}}{(2\pi\bar{h})^{\frac{1}{2}} e\vec{\mathbf{E}}_{\mathbf{Li}} t_c} \exp\left(-\frac{\Delta\vec{r}}{n_{Li}}\frac{\nabla\Psi_{Li}^{e}}{\Psi_{Li}^{e}}\right) \tag{29}$$

and for the sodium ion electric field the second-order correlation function is

$$g_{Na}^{(2)} = 1 + \exp[-\pi(\alpha_{Na})^2] \tag{30}$$

$$\alpha_{Na} = \frac{\omega_{c_Na}}{\omega_{\gamma_Na}} \tag{31}$$

$$\omega_{c_Na} = \omega_{e_Na}\lambda_{Na} \tag{32}$$

$$\omega_{\gamma_Na} = \frac{a^3}{2\bar{h}}\left[ {n_{Na}}^2 {\vec{E}_{Na}}^2 \right] \tag{33}$$

$$\omega_{e_Na} = \frac{e\vec{\mathbf{E}}_{\mathbf{Na}}}{\bar{h}\,(3\pi^2\bar{n}_c)^{\frac{1}{3}}} \tag{34}$$

$$\lambda_{Na} = \frac{a^2 {m_{eff}}^{3/2} \omega_{e_Na}^{3/2}}{(2\pi\bar{h})^{1/2} e\vec{\mathbf{E}}_{\mathbf{Na}} t_c} \exp\left(-\frac{\Delta\vec{r}}{n_{Na}}\frac{\nabla\Psi_{Na}^{e}}{\Psi_{Na}^{e}}\right) \tag{35}$$

In equations 25 and 31, $\alpha_{Li}$ and $\alpha_{Na}$ are defined as the ratio of the lithium or sodium coherent angular frequency $\omega_{c_Li}$ or $\omega_{c_Na}$ to the lithium or sodium total electric angular frequency $\omega_{\gamma_Li}$ or $\omega_{\gamma_Na}$ respectively. Equations 28 and 34 are the angular frequencies $\omega_{e_Li}$ and $\omega_{e_Na}$ for electron particles in $Li^{+}$/c-Si and electron waves in $Na^{+}$/a-Si respectively. The lambda functions $\lambda_{Li}$and $\lambda_{Na}$for lithiated and sodiated silicon nanowires are stated in equations 29 and 35 which were derived from quantum mechanical path integral method [14]. The lambda functions of $\lambda_{Li}$and $\lambda_{Na}$ are Gaussian equations that are dependent on the transition state vector $\Delta\vec{r}$ which is defined as the transitional length from an initial state to a final state of a wavefunction. The excited state wavefunction for lithium ions $\Psi_{Li}^{e}$ and sodium ions$\Psi_{Na}^{e}$ are constructed by using time-independent perturbation theory, the coherence time is $t_c$, and the refractive indices are $n_{Li}$ and $n_{Na}$ for lithium and sodium respectively.

The relationships between second-order correlation functions $g_{Li}^{(2)}$ and $g_{Na}^{(2)}$are displayed in figure 4. In general, when $g^{(2)}(\omega) = 1$ the electric field is in a coherent optical state and conversely, when $g^{(2)}(\omega) = 2$ the electric field will be in an incoherent optical state. The mixed optical state is defined as $1 < g^{(2)}(\omega) < 2$. When the second-order correlation functions is in the mixed optical state for lithiated silicon, the shape of $g_{Li}^{(2)}$ is an inverted Gaussian function where the low point of the function is $g_{Li}^{(2)} = 1.1766$.

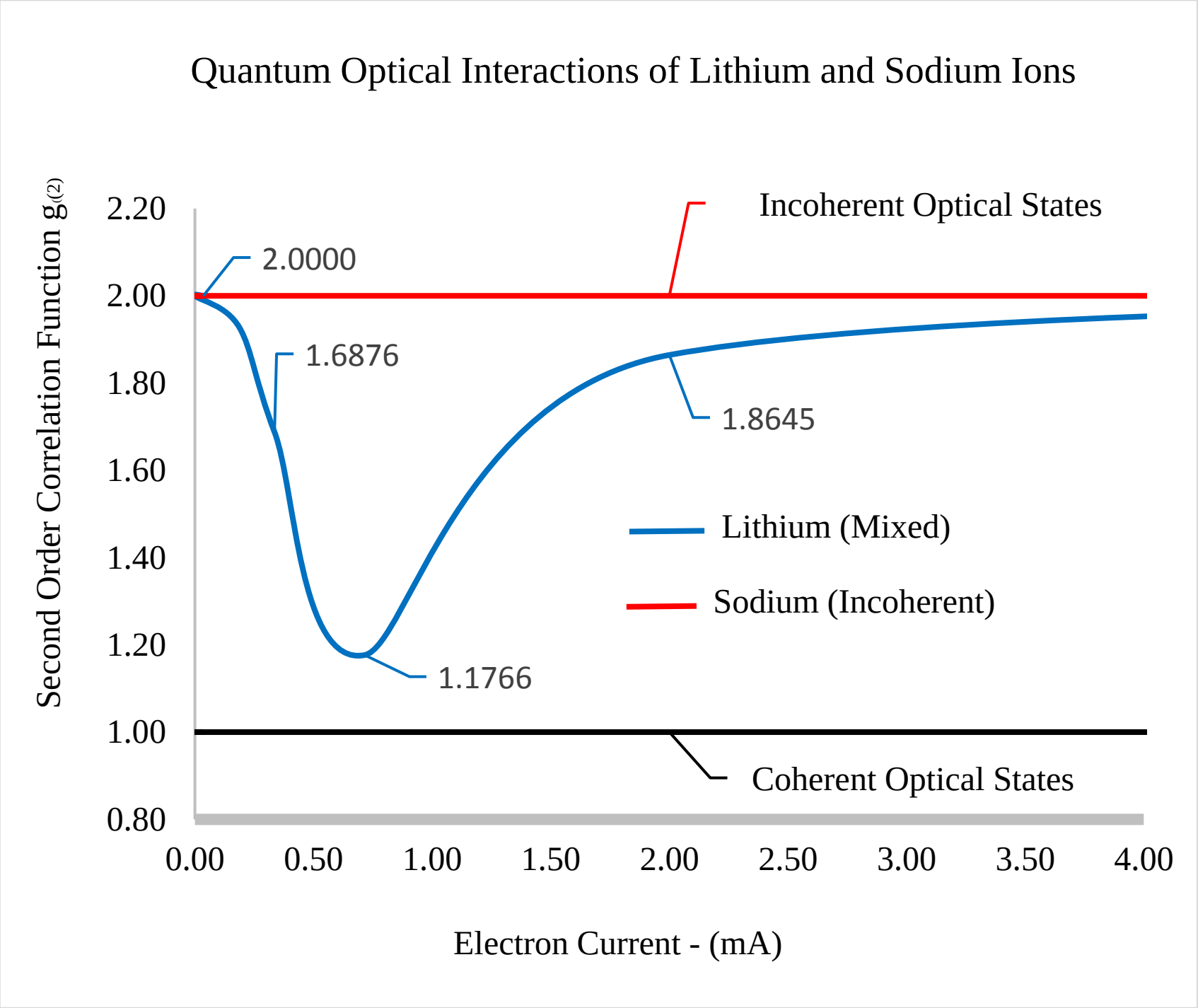

**Figure 4.** The second-order correlation functions $g^{(2)}$for lithiated and sodiated silicon is displayed. Lithium is in the mixed optical state since $1 < g_{Li}^{(2)} < 2$, however $g_{Na}^{(2)} \approx 2$ and therefore is define as being in the incoherent optical state.

## 4. Specific Charge Capacity

The coherent optical interactions that are produced from the electric fields during optical amplification is directly proportional to specific charge capacity (SCC) of the lithium and sodium silicon nanowires

$$SCC_{Li} = (1 + \epsilon_{Li})\frac{\omega_{c_Li}}{\omega_{Li}}\frac{\overline{N}_c}{\overline{N}_{Li}}\frac{e}{m_{Li}} \quad (36)$$

$$SCC_{Na} = (1 + \epsilon_{Na}) \frac{\omega_{c_Na}}{\omega_{Na}} \frac{\overline{N}_c}{\overline{N}_{Na}} \frac{e}{m_{Na}} \quad (37)$$

where $\epsilon_{Li}$ and $\epsilon_{Na}$ are the volumetric strain for lithiated and sodiated silicon nanowires respectively. The volumetric stains are constant in this study because both silicon nanowires are at their maximum volume during the computational analysis [15]. The minimum $g^{(2)}_{Li}$ is related to the maximum SCC for lithium as shown in figures 4 and 5 respectively. The second-order correlation function $g^{(2)}_{Li}$ is inversely proportional to specific charge capacity $SCC_{Li}$.The result is that the coherent optical state within the lithiated silicon nanowire increases due to the optical amplification process which leads to an increase in lithium specific charge capacity $SCC_{Li}$. Since optical amplification does not occur within the sodiated silicon nanowire, the Gaussian function $\lambda_{Na}$ for sodium is 'flat' and as a result the sodium $SCC_{Na}$ is significantly lower compared to lithium $SCC_{Li}$ (figure 5).

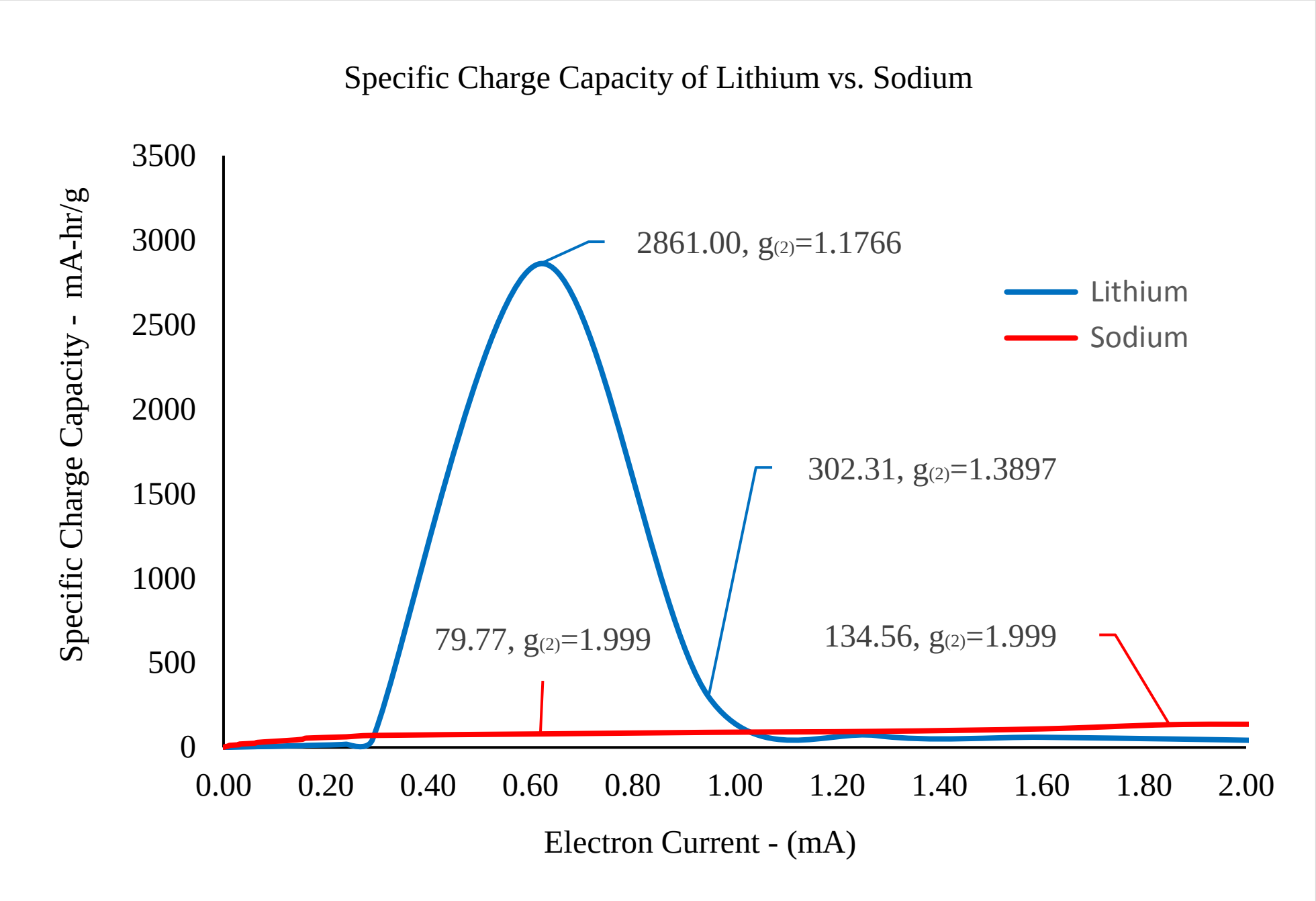

**Figure 5.** Comparing the specific charge capacity of lithium ion versus sodium ion diffusion into silicon nanowires. Lithiated silicon has a large surge of energy at approximately 0.60 milliamps which manifest into a $SCC_{Li} = 2861$ mA-hr/g due to optical amplification that develops a majority of coherent optical interactions. However, sodiated silicon with optical amplifications approximately zero develops mostly incoherent optical interactions and as a result low amounts of specific charge capacity ($SCC_{Na} = 79.77$ mA-hr/g at 0.60 mA).

## 5. Summary

In this research study we have examine quantum mechanical properties of lithium and sodium ion insertion into crystallized and amorphous silicon nanowires respectively with an electron current flowing in an opposing direction from the ion diffusing. It has been demonstrated through computational analysis that lithiated silicon can generated a large electric field $\vec{\mathbf{E}}_{\mathbf{Li}}$ that includes optical amplification. Conversely, the computational analysis predicts that the electric field $\vec{\mathbf{E}}_{\mathbf{Na}}$ inside the sodiated silicon nanowire is extremely weak with no optical amplification. With the vastly different electric fields magnitudes in each silicon nanowire, the electron current in lithiated silicon was modeled as electron particles from a large coherent energy state system and the sodiated silicon electron current was modeled as electron waves due the potential energy was calculated to be approximately zero. The optical amplification is reason for the high levels of specific charge capacity $SCC_{Li}$ in lithiated silicon through the process of coherent optical interaction which is generated by electric field $\vec{E}_{Li}$. Since optical amplification does not develop in sodiated silicon, the incoherent optical interactions are predominately present in sodiated silicon and an extremely low levels of $SCC_{Na}$ is present in sodium ion silicon nanowires.